\documentclass[runningheads]{llncs}
\usepackage{graphicx}
\usepackage{multirow}
\usepackage{subcaption}
 \usepackage{amsmath} 
 \usepackage{amssymb}
 \usepackage[svgnames]{xcolor}
 \usepackage{tabularx}
 \usepackage[misc,geometry]{ifsym}
 
\begin{document}
\title{Toward Automatic Group Membership Annotation for Group Fairness Evaluation \thanks{Supported by Institute for Financial Services Analytics at the University of Delaware}}
\titlerunning{Toward Automatic Group Membership Annotation}
%
\author{Fumian Chen \Letter\inst{1}\orcidID{0009-0001-2391-6578} \and
Dayu Yang\inst{2}\orcidID{0009-0006-7360-1837} \and
Hui Fang\inst{3}\orcidID{0009-0003-1904-787X}}
\authorrunning{F. Chen et al.}
%

\institute{University of Delaware, Newark DE 19702, USA\inst{1,}\inst{2,}\inst{3} \\
\email{\{fmchen,dayu,hfang\}@udel.edu}} 

\maketitle              
\begin{abstract}
With the increasing research attention on fairness in information retrieval systems, more and more fairness-aware algorithms have been proposed to ensure fairness for a sustainable and healthy retrieval ecosystem. However, as the most adopted measurement of fairness-aware algorithms, group fairness evaluation metrics, require group membership information that needs massive human annotations and is barely available for general information retrieval datasets. This data sparsity significantly impedes the development of fairness-aware information retrieval studies. Hence, a practical, scalable, low-cost group membership annotation method is needed to assist or replace human annotations. This study explored how to leverage language models to automatically annotate group membership for group fairness evaluations, focusing on annotation accuracy and its impact. Our experimental results show that BERT-based models outperformed state-of-the-art large language models, including GPT and Mistral, achieving promising annotation accuracy with minimal supervision in recent fair-ranking datasets. Our impact-oriented evaluations reveal that minimal annotation error will not degrade the effectiveness and robustness of group fairness evaluation. The proposed annotation method reduces tremendous human efforts and expands the frontier of fairness-aware studies to more datasets.

\end{abstract}


\section{Introduction}
\vspace{-2mm}
From social media to open web searches, information retrieval (IR) systems are ubiquitous and can fundamentally impact how people receive and seek information. As people started to notice the issue of the echo chamber, the polarized online community, and the importance of covering diverse results \cite{ekstrand2019fairness}, fairness-aware IR and its evaluation metrics became emerging needs to combat unfairness and biased representation for long-term sustainability \cite{zehlike2021fairness}. Group fairness evaluation metrics are the most adopted metrics, measuring the disparity between a situation to be evaluated and its ideal situation. When applying them, one of the necessities is the group membership (GM) annotations, which define whether an item is from underrepresented groups. Without GM annotation, applying group fairness evaluation metrics on retrieval results is infeasible, and it is also impossible to apply supervised learning-based fair ranking algorithms that rely on GM annotation for training \cite{chen2023learn,zehlike2020reducing}. Therefore, before evaluating group fairness or allocating exposure to the documents, we must know their group membership.

Annotations are usually obtained through costly human annotators, such as crowd annotators and domain experts. High-quality annotations involving annotators' training,  and cross-validation are even more expensive \cite{kasthuriarachchy2021cost}. The annotation process requires annotators to interpret documents' context and then assign pre-defined labels to the documents based on their contextual information. Since it is very similar to a text classification process, various attempts have been proposed to assist or replace human annotations, especially with the emergence of advanced NLP techniques that can accurately capture contextual features from text \cite{ishita2020using,kasthuriarachchy2021cost}. However, most of these attempts to replace human annotations focus on accuracy compared with human annotation but ignore the impact when enforcing this replacement on different tasks. It remains unclear how annotation errors would impact the final metrics with machine-learned annotations, especially when previous studies have shown that document-level error might be eliminated when aggregating to higher levels \cite{bailey2008relevance}. Since group fairness evaluations are also aggregated metrics, the annotation error might not hurt the ability to evaluate fairness for IR systems. The relation between annotation accuracy and the final evaluation metrics deserves our attention. Moreover, even though generative large language models (LLMs) are not designed for discriminative tasks like text classification, the increasing trend of using generative large language models (LLMs) such as OpenAI GPT on downstream NLP tasks is pushing more and more researchers to scramble for their applications \cite{goel2023llms}. However, given its economical and computational cost, are generative models with billions of parameters better than discriminative models for fairness-related annotation tasks? 

Therefore, to explore how to replace human GM annotation effectively and economically and solve the issue of data sparsity, we compared the performance of four representative language models in predicting group membership for group fairness evaluation. Then, we comprehensively studied the impact of replacing human GM annotation for group fairness evaluations in recent fair-ranking datasets. Confirming the effectiveness of the new GM annotation method with minimal supervision, we believe our work opened a new direction to reduce human efforts on GM annotation and augment traditional IR datasets for future fairness-aware studies. Our implementation code will be available at \url{https://github.com/fm-chen/nldb-experiments}.

\vspace{-4mm}
\section{Related Work}
\vspace{-2mm}
With the rapid development of NLP, especially with the emergence of masked language models such as BERT \cite{devlin2018bert} and generative large language models like OpenAI GPT \cite{ray2023chatgpt}, more and more NLP-related work has been proposed to save or even replace human efforts. Text classification, which assigns one of the pre-defined labels to a given text sequence, is one of the classic NLP tasks. As one of the most powerful language models, BERT provides various pre-trained models that accurately capture linguistic and semantic information out of text \cite{koroteev2021bert}. With proper fine-tuning, previous studies have used BERT for multiple annotation tasks, such as image labeling and dataset annotation, and shown promising results even with fewer training samples and imbalanced class distributions \cite{ma2023effectiveness,laurer2022less}. Recently, as generative LLMs have become a hot topic, OpenAI GPT has also attracted increasing research interest, including the use of GPT to assist annotation and labeling. Generative models like GPT have shown to be a valuable tool for predicting searcher preference, validating and assisting human annotations and labelings \cite{thomas2023large,pangakis2023automated,he2023annollm,ding2022gpt}. Compared with BERT, which has a parameter size from 30 million to about 350 million, LLMs usually involves billions to over hundred billions of parameters, making fine-tuning and using LLMs costly \cite{ding2022gpt}. Even with the open-sourced LLM, Mistral with seven billion parameters \cite{jiang2023mistral}, deploying the model locally is computationally costly. Since generative models are not designed for discriminative tasks like text classification, previous studies revealed that using LLMs effectively requires meticulously prompt design. Their performance varies dramatically under different contexts \cite{chae2023large,zhang2024pushing}. Therefore, instead of scrambling for LLMs, we would like to explore the accuracy and impact of using different language models to replace human GM annotations for fairness evaluation tasks.

This work is also closely related to fairness-aware IR and its evaluation metrics. Well-adopted fairness evaluations \cite{diaz2020evaluating,gao2022fair,sapiezynski2019quantifying,singh2018fairness,raj2020comparing} were based on exposure, and their fairness metrics either measure the deviation between system-produced and target exposure distribution or measure the inequality of exposure across groups. Another group of fairness evaluation is based on pair-wise metrics measuring the difference between pairs \cite{beutel2019fairness,narasimhan2020pairwise}. They are all aggregated measures that treat groups instead of individual documents as the basic unit, and the impact of replacing the costly human GM annotation with NLP techniques is unclear and has never been studied before. Thus, to save human efforts in obtaining GM annotations and solve the issue of data sparsity in fairness evaluation, this study tested four language models to obtain GM annotations and explored the impact of replacing human annotations with different annotators.

\begin{figure}[hbt!]
  \vspace{-4mm}
      \centering
      \includegraphics[width=\linewidth]{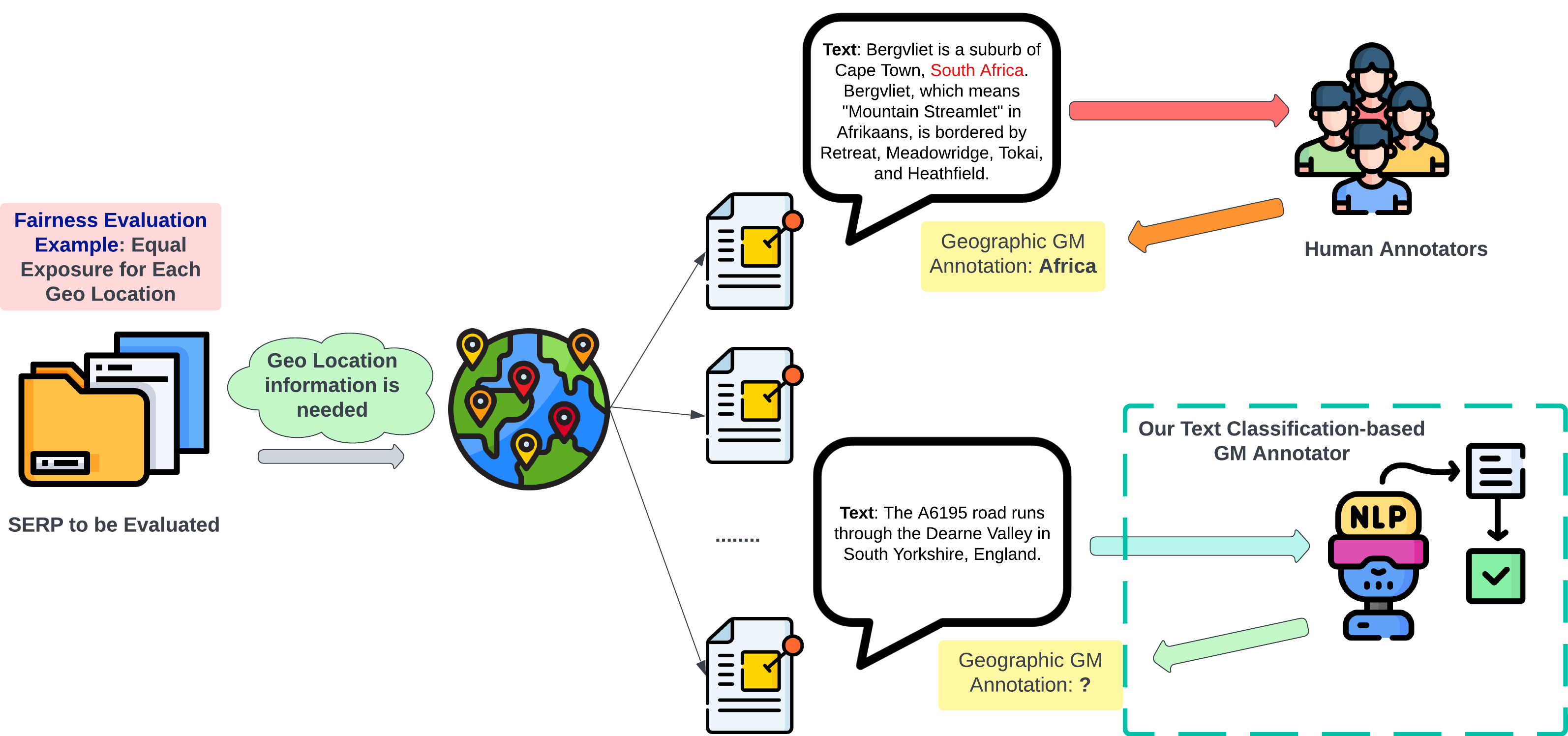}
      \caption{The necessity of GM annotation in group fairness evaluation}
      \label{fig:gen-flow}
      \vspace{-5mm}
  \end{figure}

\vspace{-4mm}
\section{Automate GM Annotation for Fairness Evaluation}
\vspace{-2mm}
\subsection{GM in Fairness Evaluations}
\vspace{-2mm}
Group membership (GM) is one of the most essential components in group fairness evaluation. Depending on fairness evaluation goals, group membership can involve one or more fairness categories, such as gender and geographic location. As shown in Fig. \ref{fig:gen-flow}, to make sure that a search engine result page (SERP) contains items from different geographic locations, we have to know each item's geographic location information (geographic GM annotation) first. With the GM annotation, merits or exposure distributions across groups can be formulated to construct fairness evaluation. For example, the TREC fair ranking track 2021 \cite{trec-fair-ranking-2021} and 2022 \cite{ekstrand2023overview} \footnote{\url{https://fair-trec.github.io/}} use the attention-weighted rank fairness (AWRF), a widely used exposure-based fairness evaluation measuring the difference between ranking $L$'s cumulative exposure $\epsilon_{L}$ and population estimator $\hat{\epsilon}$:  \vspace{-1mm}
\begin{equation*}
    \text{AWRF(L)} = \Delta(\epsilon_{L}, \hat{\epsilon}) 
\end{equation*}
where $\Delta$ is a divergence function (e.g., Kullback–Leibler divergence or Jenson-Shannon divergence). The ranking $L$'s cumulative exposure $\epsilon_{L}$ is computed by $\sum_{d \in L} w(L) * GM_{d}$ where $w(L)$ is an attention decay function and $GM_{d}$ is the group membership matrix of document $d$. For instance, $GM_{d} (\text{Gender}) = (1, 0, 0)$ if the document $d$ is annotated as group ``male" for fairness category gender with three subgroups: ``male", ``female", and ``non-binary". The population estimator $\hat{\epsilon}$ reflects the target exposure distribution that a fair system should produce, which could also rely on GM annotation. TREC estimates $\hat{\epsilon}$ by averaging the group membership of all relevant documents to ensure that each group of items receives the same amount of expected exposure as their relevance grade. Moreover, the target exposure distribution can also be given. For example, the NTCIR fairweb1 task \cite{tao2023overview} assumes a uniform distribution across groups as their target.

\begin{table}[hbt!]
  \vspace{-4mm}
  \centering
  \caption{Task description and fairness categories: Internal fairness categories are internal attributes which do not require human annotation. We focus on contextual fairness categories.}
  \resizebox{\textwidth}{!}{%
  \begin{tabular}{c|c|cc}
  \hline\hline
  \multirow{2}{*}{\textbf{Dataset}} &
    \multirow{2}{*}{\textbf{\begin{tabular}[c]{@{}c@{}}Task\\ Description\end{tabular}}} &
    \multicolumn{2}{c}{\textbf{Fairness Categories}} \\ \cline{3-4} 
   &
     &
    \multicolumn{1}{c|}{\textbf{Contextual}} &
    \textbf{Internal} \\ \hline \hline
  \begin{tabular}[c]{@{}c@{}}TREC fair ranking\\ track 2021\end{tabular} &
    \multirow{2}{*}{\begin{tabular}[c]{@{}c@{}}A Wikipedia article fair ranking task\\ (corpus containing more than 6 million \\ articles): provide fair exposure for each \\ group of  documents regarding \\ different fairness categories.\end{tabular}} &
    \multicolumn{1}{c|}{\begin{tabular}[c]{@{}c@{}}(1) Gender of article's \\ subject (Gender, 4 subgroups)\\  (2) Geographical location \\ associated with the article \\(Geo, 8 subgroups)\end{tabular}} &
    N/A \\ \cline{1-1} \cline{3-4} 
  \begin{tabular}[c]{@{}c@{}}TREC fair ranking\\ track 2022\end{tabular} &
     &
    \multicolumn{1}{c|}{\begin{tabular}[c]{@{}c@{}}(1) Gender of article's \\ subject (Gender, 4 subgroups)\\ (2) Geographical location \\ associated with the article \\(Geo, 21 subgroups)\end{tabular}} &
    \begin{tabular}[c]{@{}c@{}}(1) Age of the article\\ (2) Occupation\\ (3) Alphabetical orders\\ (4) Popularity (5) Replication \\ in other languages\end{tabular} \\ \hline
  NTCIR fairweb1 &
    \begin{tabular}[c]{@{}c@{}}A fair ranking tasks (corpus \\ Chuweb21D containing about \\ 50 million documents): provide \\ group-fair results for research,\\ movie, and YouTube content.\end{tabular} &
    \multicolumn{1}{c|}{\begin{tabular}[c]{@{}c@{}} (1) Movie's country of origin \\ (Movie-Origin, 8 subgroups)\\(2) Gender of researcher\\ (Research-Gender, 3 subgroups)\end{tabular}} &
    \begin{tabular}[c]{@{}c@{}}(1) Research-Hindex\\ (2) Movie-Ratings\\ (3) YouTube-Subscription\end{tabular} \\ \hline \hline
  \end{tabular}%
  }
  \vspace{-2mm}
  \label{tab:fair-cate}
  \vspace{-4mm}
  \end{table}

\vspace{-3mm}
\subsection{Challenges with GM annotation}
\begin{figure}[]
  \vspace{-4mm}
      \centering
      \includegraphics[width=0.8\linewidth]{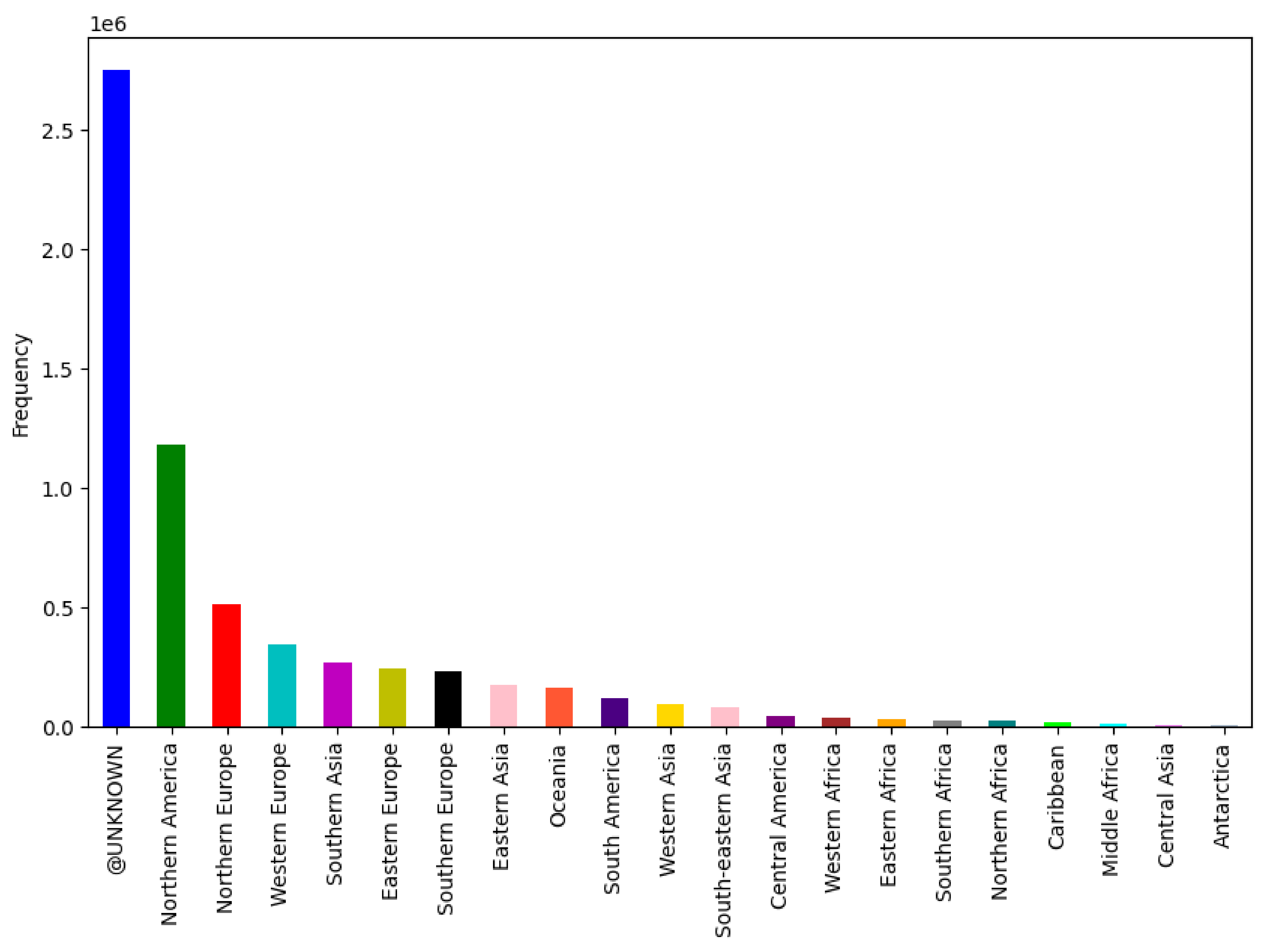}
      \caption{Geo subgroup frequency of human GM annotation (TREC 2022).}
      \label{fig:freq_2022}
      \vspace{-5mm}
  \end{figure}

\vspace{-1mm}
Obtaining GM annotation can be challenging and requires significant human effort. We investigate three recent fair-ranking tasks: (1) TREC fair ranking track 2021 \cite{trec-fair-ranking-2021}, (2) TREC fair ranking track 2022 \cite{ekstrand2023overview}, and (3) the NTCIR Fair Web task \cite{tao2023overview}. Details about these tasks are reported in Table \ref{tab:fair-cate}. TREC fair ranking tasks are based on a Wikipedia corpus containing more than six million English articles and 50/50 training and evaluating queries from various domains, whereas NCTIR fairweb1 is based on an English document collection, Chuweb-21D, containing more than 40 million documents, including research papers, movies, and YouTube Content. Unlike many previous fair-ranking studies based on outdated datasets that only contain numeric features, all three tasks provide full-text fields and enable us to apply NLP techniques. They also offer page meta information consisting of human annotations. Fig. \ref{fig:freq_2022} shows the subgroups' frequency of human annotation by page geographic locations in the TREC 2022 datasets. As can be seen, the documents' geographic information was annotated into 21 subgroups, and a huge imbalance exists across groups. Almost half of the documents were marked as ``unknown" because they either lacked annotation or were non-applicable. Ensuring a high-quality annotation is challenging, given inevitable human error and costly knowledge training for human annotators, let alone annotating GM into large numbers of subgroups. Given these challenges, few datasets with GM annotation are available for fairness-aware studies. Therefore, we aim to automate GM annotation with minimal human efforts to break the data sparsity using NLP techniques.

\vspace{-4mm}
\subsection{Annotating GM by Text Classification with Language Models} \label{text-models}
\vspace{- 1.5mm}
The quality of human annotation heavily depends on the annotators' knowledge and interpretation of the raw text. The annotation process is similar to text classification algorithms that capture contextual patterns (interpretation of raw text) and categorize raw text based on training data (knowledge). Accordingly, we assume that replacing human annotation with text classification models is possible by adequately utilizing text information, especially with sophisticated language models that can precisely capture linguistic and semantic information and even outperform humans in some studies. In this work, we explored the following text classification models for GM annotation:
\vspace{-1.5mm}
\begin{itemize}
    \item \textbf{Linear BoW Model}: a linear bag-of-words model \cite{zhang2010understanding} followed by a neural network classifier implemented by \textit{spaCy} \textit{TextCategorizer}\footnote{\url{https://spacy.io/api/textcategorizer}}.
    \item \textbf{BERT-based Model}: a fine-tuned BERT sentence classification model \cite{devlin2018bert} \textit{"bert-large-uncased"} \footnote{\url{https://huggingface.co/bert-large-uncased}} implemented by \textit{PyTorch} \footnote{\url{https://pytorch.org/}}. 
    \item \textbf{GPT Models}: a generative large language model, with GPT-3.5-turbo and GPT-4, implemented by \textit{spaCy-LLM} \footnote{\url{https://spacy.io/usage/large-language-models}}.
    \item \textbf{Mistral 7B Models \cite{jiang2023mistral}}: a generative large language model, Mistral-7B-Instruct-V0.2 \footnote{\url{https://huggingface.co/mistralai/Mistral-7B-Instruct-v0.2}}, implemented by \textit{PyTorch}.
\end{itemize}
\vspace{-1.5mm}
Linear bag-of-words (BoW) model \cite{zhang2010understanding} is one of the simplest statistical language models (SLM) that convert words to numeric representations based on vocabulary set and word count. It is flexible and performs well for simple document classification tasks, but it cannot understand context. Bidirectional Encoder Representations from Transformers (BERT) \cite{devlin2018bert}, introduced in 2018, is one of the masked language models (MLM) that can successfully capture semantic and linguistic information from text sequences, which has dominated classification tasks since being introduced. In contrast, generative large language models, such as OpenAI GPT, are not designed for classification tasks but have shown potential in assisting human annotations in recent studies \cite{goel2023llms}. Unlike GPT, which is fully commercialized and expensive, Mistral 7B \cite{jiang2023mistral} is a state-of-the-art, open-sourced LLM that achieved promising performance across various benchmark tasks. The performance of generative LLMs varies task by task and is heavily dependent on their pre-trained data and prompt design. Given the advantages and limitations of these language models, we would like to explore their capability of replacing GM annotation for group fairness evaluations.

\begin{table}[]
  \centering
  \caption{LLM prompts. Shaded text is optional for one-shot or fine-tuning.}
  \label{tab:prompt}
  \resizebox{\textwidth}{!}{%
  \begin{tabular}{p{0.2\linewidth}|p{0.8\linewidth}}
  \hline \hline
  \multicolumn{1}{c|}{\textbf{Model}} &
    \multicolumn{1}{c}{\textbf{Prompt}} \\ \hline \hline
  \textbf{GPT-3.5-turbo/GPT-4}&
    \begin{tabular}[c]{@{}l@{}}You are an expert Text Classification system. \\ Your task is to accept text as input and provide a category for \\ the text based on the pre-defined labels. Classify the text below to \\ any of the following labels: {[}GM Labels{]} \colorbox{lightgray}{Below are some} \\ \colorbox{lightgray}{examples (only use these as a guide): {[}Example Text{]}, {[}Answer{]}.}\\ Here is the text that needs classification: {[}Text{]}\end{tabular} \\ \hline
  \textbf{Mistral} &
    \begin{tabular}[c]{@{}l@{}}{[}INST{]}Analyze the {[}Fairness Category{]} of the Wikipedia article\\  enclosed in square brackets,  determine if it is {[}GM Labels{]}, \\ and return the answer as the corresponding labels {[}/INST{]} \\ \colorbox{lightgray}{{[}Example Text{]} = {[}Answer{]}}\end{tabular} \\ \hline \hline
  \end{tabular}%
  }
  \vspace{-5mm}
  \end{table}

We build classification models trained or fine-tuned by small-size human-annotated samples using these language models for GM annotation. For each subgroup within a fairness category, we equally sampled 500 training documents and 100 testing documents from each group. We follow the standard data cleaning process for the text field, including special character removal, stop word removal, and lemmatization. Given the average length of Wikipedia articles, 670 tokens, we truncated the full-text field to 512 tokens without losing much information. The optimal model weights are trained or fine-tuned on training samples and obtained by minimizing a KL-divergence classification loss for the linear-bag-of-words and BERT-based models. For the GPT models, we use the prompt shown in Table \ref{tab:prompt} provided by \textit{spaCy} and set the template to 0.3 for one-shot text classification. To use the Mistral models (\textit{Mistral-7B-Instruct-V0.2}), we utilized low-rank adaption \cite{hu2021lora} so that we can computationally run the model with our best GPU. The prompt used for Mistral is also reported in \ref{tab:prompt}. Given the high cost of fine-tuning GPT models, we only fine-tuned the Mistral 7B in this study. Finally, we use these text classifiers as annotators to predict the group membership information of new documents. Once fairness annotations are obtained, ideally, we can fit them into any group fairness evaluation metrics or augment other IR datasets for fairness-aware studies.

\vspace{-2mm}
\section{Evaluation and Analysis}
\vspace{-2mm}

\newcolumntype{C}{>{\centering\arraybackslash}X}

\begin{table}[]
\vspace{-4mm}
  \centering
  \caption{Classification performance (accuracy and f-1 scores) by different models when predicting ``Gender" group membership: ``male", ``female", ``non-binary" and ``unknown" (TREC 2022). The Mistral model is fine-tuned with a full-, partial-, and proportional- set of training examples. * indicates the best-performed model.}
  \label{tab:cl-perform}
  \begin{tabularx}{\textwidth}{l|C|C|C|C|C|C}
    \hline \hline
    \textbf{Models} & \textbf{\begin{tabular}[c]{@{}c@{}}Overall\\ Accuracy\end{tabular}} & \textbf{\begin{tabular}[c]{@{}c@{}}Overall\\ F-1\end{tabular}} &  \textbf{\begin{tabular}[c]{@{}c@{}}Male\\ F-1\end{tabular}} &  \textbf{\begin{tabular}[c]{@{}c@{}}Female\\ F-1\end{tabular}} &  \textbf{\begin{tabular}[c]{@{}c@{}}NB\\ F-1\end{tabular}} &  \textbf{\begin{tabular}[c]{@{}c@{}}Unknown\\ F-1\end{tabular}} \\ 
    \hline \hline
    \textbf{Linear-BoW} & 0.905 & 0.9073 & 0.8475 & 0.9800 & 0.8889 & 0.9130 \\
    \hline
    \textbf{BERT*} & 0.985* & 0.9850* & 0.9804* & 0.9899* & 0.9697* & 1* \\
    \hline
    \textbf{GPT-3.5-turbo} & 0.820 & 0.7947 & 0.8727 & 0.9009 & 0.5507 & 0.8545 \\
    
    \textbf{GPT-4} & 0.865 & 0.8549 & 0.8403 & 0.9259 & 0.6842 & 0.9691 \\
    \hline
    \textbf{Mistral (zero-)} & 0.655 & 0.6446 & 0.6076 & 0.6565 & 0.5915 & 0.7227 \\
    
    \textbf{Mistral (full-)} & 0.705 & 0.6921 & 0.7912 & 0.7458 & 0.5135 & 0.7179 \\
    
    \textbf{Mistral (part-)} & 0.425 & 0.3564 & 0.6619 & 0.1515 & 0.4754 & 0.1370 \\
    
    \textbf{Mistral (prop-)} & 0.655 & 0.6624 & 0.8211 & 0.5686 & 0.5487 & 0.7111 \\
    \hline \hline
  \end{tabularx}
  \vspace{-1mm}
\end{table}

\subsection{Prediction Accuracy of GM Annotation Models}
\vspace{-1mm}

\begin{figure}[hbt!]
  \centering
  \includegraphics[width=0.7\linewidth]{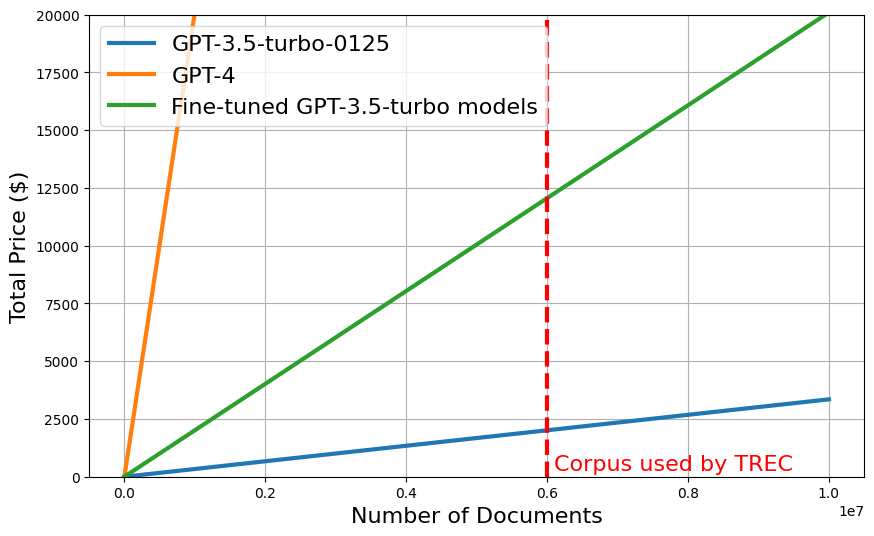}
  \caption{Total price of annotation using trained GPT models (GPT-4, GPT-3.5-turbo, and fine-tuned GPT-3.5) by number of documents.}
  \label{fig:gpt-price}
  \vspace{-6mm}
\end{figure}

We first examine the annotation accuracy between these annotation models when predicting gender GM annotation. The performance of each classifier is reported in Table \ref{tab:cl-perform}. As can be seen, generative models (LLMs) failed to outperform the discriminative models (BERT and BoW models), especially for the gender subgroup ``non-binary." This might be because LLMs were pre-trained on biased data where the subgroup ``non-binary" was under-represented, which is currently a known issue \cite{lucy2021gender}. If we do not want to amplify this pre-existing bias, fine-tuning LLM-based models is required. Given the long text length and large corpus size (e.g., about 6 million for the TREC fair ranking track) to annotate, fine-tuning GPT models and generating GM annotations for new documents would be extremely expensive. The total price of annotating datasets with a size similar to the TREC corpus is over \$2000 even with the cheapest GPT-3.5-turbo model, as shown in Fig. \ref{fig:gpt-price}. With the open-sourced LLM, Mistral, its fine-tuned models still cannot correctly predict the label of ``non-binary", including using different fine-tuning strategies to improve its performance as shown in the last four rows in Table \ref{tab:cl-perform}. Since Mistral has difficulty to predict subgroup ``Male" and ``non-binary" correctly, we first fine-tuned Mistral with ``Male" and ``non-binary" only but as shown in the Table \ref{tab:cl-perform}, we seem to have over-corrected the model. It is also the case when we fine-tuned the model with more ``Male" and ``non-binary" than the other two groups. In either case, we damage the performance of Mistral compared with the equally and fully sampled fine-tuning. The performance of GPT and Mistral shows the disadvantage of LLMs for classification tasks. Therefore, in terms of using text classification for fairness GM annotation, BERT-based models outperformed LLMs, both economically and computationally.

\begin{figure}[hbt!]
  \vspace{-2mm}
    \centering
   \begin{subfigure}[]{\textwidth}
      \includegraphics[width=\textwidth]{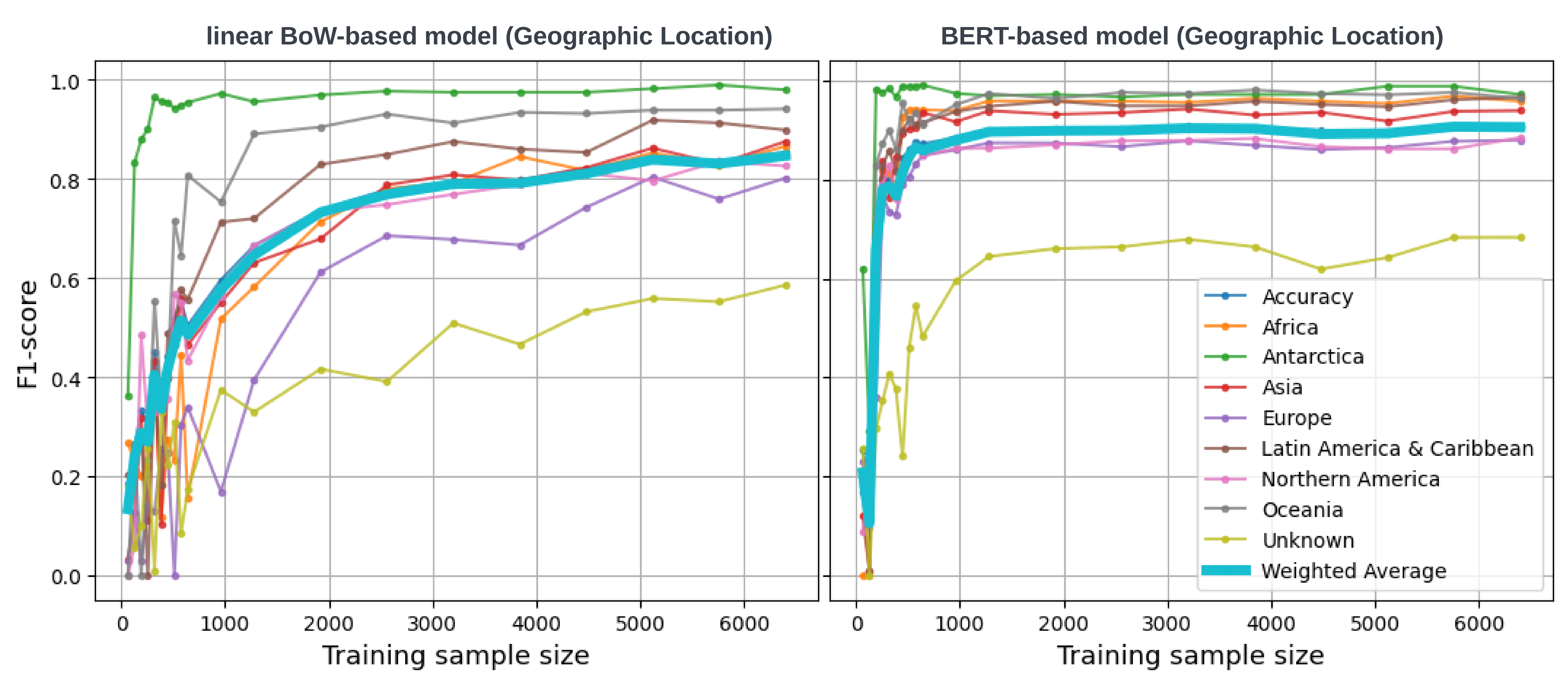}
    \end{subfigure}
    \vspace{-3mm}
    \begin{subfigure}[]{\textwidth}
      \includegraphics[width=\textwidth]{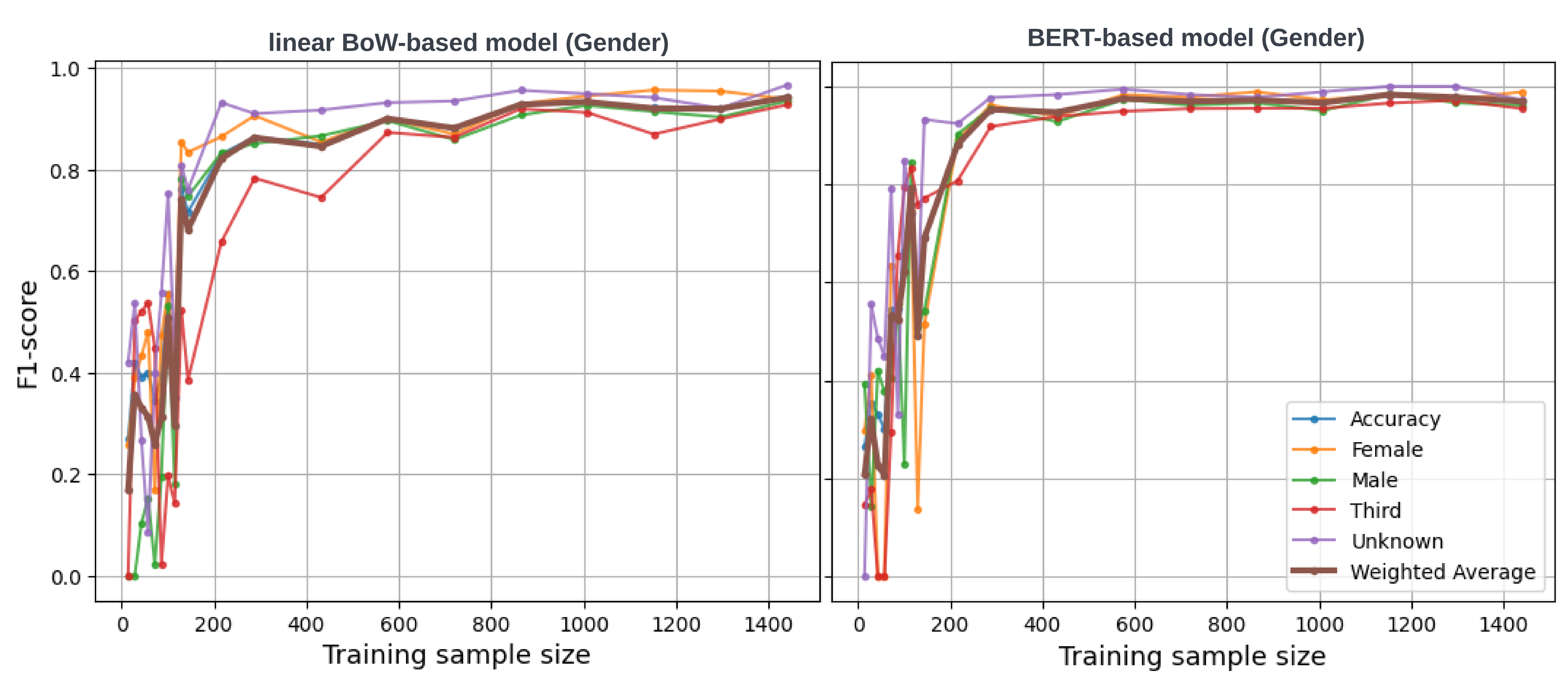}
    \end{subfigure}
    \vspace{-1mm}
    \caption{Classification performance by training sample size (TREC 2021).}
    \label{fig:q-level-pear}
    \vspace{-6mm}
\end{figure}

\begin{figure}[hbt!]
  \vspace{-6mm}
    \centering
    \includegraphics[width=\linewidth]{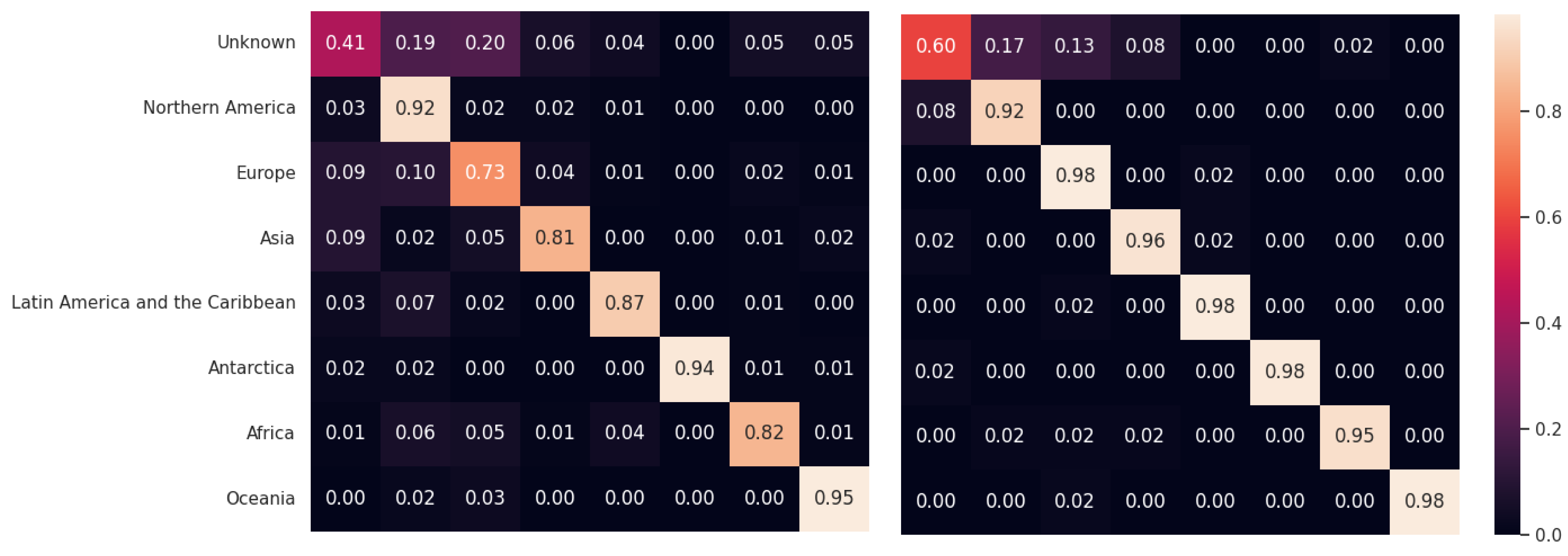}
    \caption{BERT model (right) outperformed and is less sensitive to imbalanced classes than the bag-of-words model (left) (TREC 2021).}
    \label{fig:sensitive-compare}
    \vspace{-6mm}
\end{figure}

The fine-tuned BERT-based models demonstrated a promising annotation capability and achieved the highest accuracy and f-1 scores among all models when predicting the GM annotations (It is also true for all contextual fairness categories; we only show the result for gender here to save space). This shows the advantages of the BERT sentence classification model in terms of text understanding and capturing linguistic and semantic information compared with the bag-of-words models. For both BoW-based and BERT-based models, training sample size impacts the classification performance. Fig. \ref{fig:q-level-pear} shows the text classification performance by sample size (reported as F1 scores) using both models to predict GM annotation for the TREC fair ranking track 2021. As can be seen, the linear BoW-based model requires more training samples to converge to the best performance than the BERT-based model, especially when the fairness category contains more subgroups. As shown in Fig. \ref{fig:sensitive-compare}, our results regarding geographic location GM also align with previous studies that BERT-based classifiers are less sensitive to imbalanced classes \cite{laurer2022less}. As a pre-trained model, BERT only needs a few samples to fine-tune. Compared with the size of the entire corpus, we need approximately 1200 training samples when predicting geographic location GM (8 subgroups), and 400 samples when predicting gender GM (4 subgroups) to achieve a reasonable performance using BERT sentence classification. This observation also suggests that more training samples are needed, given a fairness category with more sub-groups. Generally speaking, we recommend using no less than 100-150 training samples per subgroup when training a BERT-based model for GM annotation, depending on the number of subgroups. We also noticed that both BERT and BoW models have difficulties in predicting ``unknown" for geographic location GM, as shown in Fig. \ref{fig:sensitive-compare}. This might result from the complexity of ``unknown", which indicates either missing annotation or non-applicable. For instance, annotating geographic locations for a mathematical proof article is not very applicable.

Overall, the BERT sentence classification model is a winner based on the above comparison, given its strong performance, simplicity, and low cost. In the following sections, we explore the impact of using our best classifier to replace human GM annotation. 

\vspace{-3mm}
\subsection{Group Fairness Evaluation with GM annotations} \label{impact-ord}
\vspace{-2mm}
We showed promising text classification performance when annotating new documents using the BERT sentence classification model. Even though annotation errors still exist, we are curious about the impact of these document-level mistakes and whether these mistakes will be washed out when aggregated into aggregated-level evaluation \cite{bailey2008relevance}, since fairness evaluation is an aggregated metric. To test the effectiveness of the BERT-based GM annotation, we use the Pearson correlation coefficient test \cite{cohen2009pearson} and Spearman's rank correlation coefficient \cite{sedgwick2014spearman} to test whether evaluation metrics with our GM annotation method can effectively differentiate the fairness quality of different systems (rankings) as the old evaluation metrics can do with human annotation. Specifically, we investigate the correlation between the official evaluation metrics with human GM annotation and those with our BERT-based GM annotation. The investigation is based on all participants' official submissions to three fair-ranking tasks. Because the official submissions are from multiple groups using different ranking algorithms, we believe the fairness scores of these runs provide the best estimation of the upper and lower bound of fairness performance. There are 13 runs for the TREC fair ranking track 2021, 27 runs for the TREC fair ranking track 2022, and 28 runs for the NTCIR fairweb1 task.  

\vspace{-4mm}

\begin{table}[]
  \vspace{-4mm}
  \caption{Summary of correlation tests between tasks' official evaluation metrics with human annotation and those with BERT-based GM annotation. * indicates statistical significance ($p<0.05$). The ``Overall" group for TREC tasks is the intersectional group of Gender and Geographic Location.}
  \label{tab:cor-res}
  \resizebox{\textwidth}{!}{%
  \begin{tabular}{c|ccc|ccc|cc}
  \hline \hline
  \multirow{2}{*}{\textbf{}} & \multicolumn{3}{c|}{\textbf{TREC 2021}}           & \multicolumn{3}{c|}{\textbf{TREC 2022}}           & \multicolumn{2}{c}{\textbf{NTCIR fairweb1}} \\ \cline{2-9} 
                             & \textbf{Overall} & \textbf{Gender} & \textbf{Geo} & \textbf{Overall} & \textbf{Gender} & \textbf{Geo} & \textbf{M-Orgin}     & \textbf{R-Gender}    \\ \hline \hline
  \textbf{Pearson}           & 0.9469*          & 0.9994*         & 0.9790*      & 0.9678*          & 0.9957*         & 0.9968*      & 0.9868*              & 0.9937*              \\ \hline
  \textbf{Spearman}          & 0.8187*          & 0.9945*         & 0.9231*      & 0.9609*          & 0.9670*         & 0.9976*      & 0.9189*              & 0.9688*              \\ \hline \hline
  \end{tabular}%
  }
  \vspace{-6mm}
  \end{table}

\subsubsection{System-level Evaluation} Our system-level evaluation is based on testing the correlation between metrics using human annotation and metrics using BERT-based annotation to see whether we can effectively replace human annotation while preserving the ability to differentiate ranking fairness. Table \ref{tab:cor-res} reports the correlation between tasks' official evaluation metrics and our text classification-based evaluation metrics regarding the three fair ranking tasks: TREC fair ranking track 2021, TREC fair ranking track 2022, and NTCIR fairweb1. Based on Pearson correlation and Spearman's ranked correlation tests, evaluation metrics with our BERT-based GM annotation strongly correlated with the official evaluation metrics with human annotation, and the correlations are statistically significant. This confirms the system-level effectiveness of using BERT classification-based GM annotation in evaluating fairness. Replacing human GM annotation with BERT-based annotation preserves the ability to differentiate fairness among different runs. Even though our text classifier cannot accurately predict some subgroups of some fairness categories (e.g., the group ``unknown'' of geographic location), when aggregating documents into a system-level evaluation, we can still differentiate rankings' fairness. Since how to deal with ``unknown'' is also a challenge for human annotators, this observation also suggests that minimal annotation error will not degrade system-level fairness evaluation, and the BERT-based annotation could be a solution for estimating ``unknown''.

\vspace{-5mm}
\subsubsection{Query-level Robustness.} 
\begin{figure}
  \vspace{-1mm}
    \centering
    \includegraphics[width=1\linewidth]{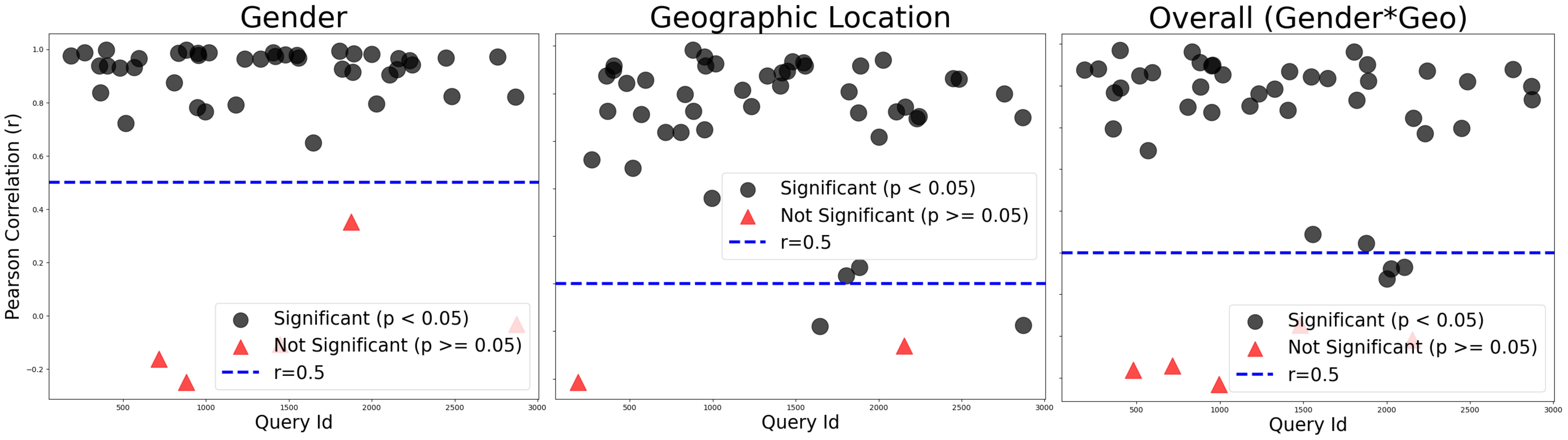}
    \caption{Query-level robustness (TREC 2022): the correlation between evaluations using human annotation and BERT-based annotation.}
    \label{fig:per-topic}
    \vspace{-4mm}
\end{figure}

The query-level evaluation decomposes the system-level evaluation by 50 evaluation queries. In Fig. \ref{fig:per-topic}, we show the Pearson correlation coefficient r between human annotation-based evaluation metrics and those using BERT-based GM annotation by the evaluation query IDs. As can be seen, for most of the queries, the correlation is high and significant. That is, we highly preserve the ability to differentiate fairness when replacing human GM annotation with BERT-based GM annotation, especially for GM of ``gender''. The ``Overall" group, which is the intersectional product of ``geographic location'' and ``gender'' also demonstrates a high correlation between human GM annotation and BERT-based annotation. Recall the Fig. \ref{fig:q-level-pear}, predicting the GM of geographic location is less accurate than predicting the GM of Gender. Therefore, with a higher accuracy of the GM annotation, we observed a more robust query-level correlation. Therefore, to be more confident in replacing human-annotated GM and evaluating at a query level, we need a text classifier that can accurately predict GM at a document level.

\vspace{-4mm}
\subsection{Impact of the Annotation Accuracy} \label{sampleSize}
\begin{figure}
    \centering
    \includegraphics[width=1\linewidth]{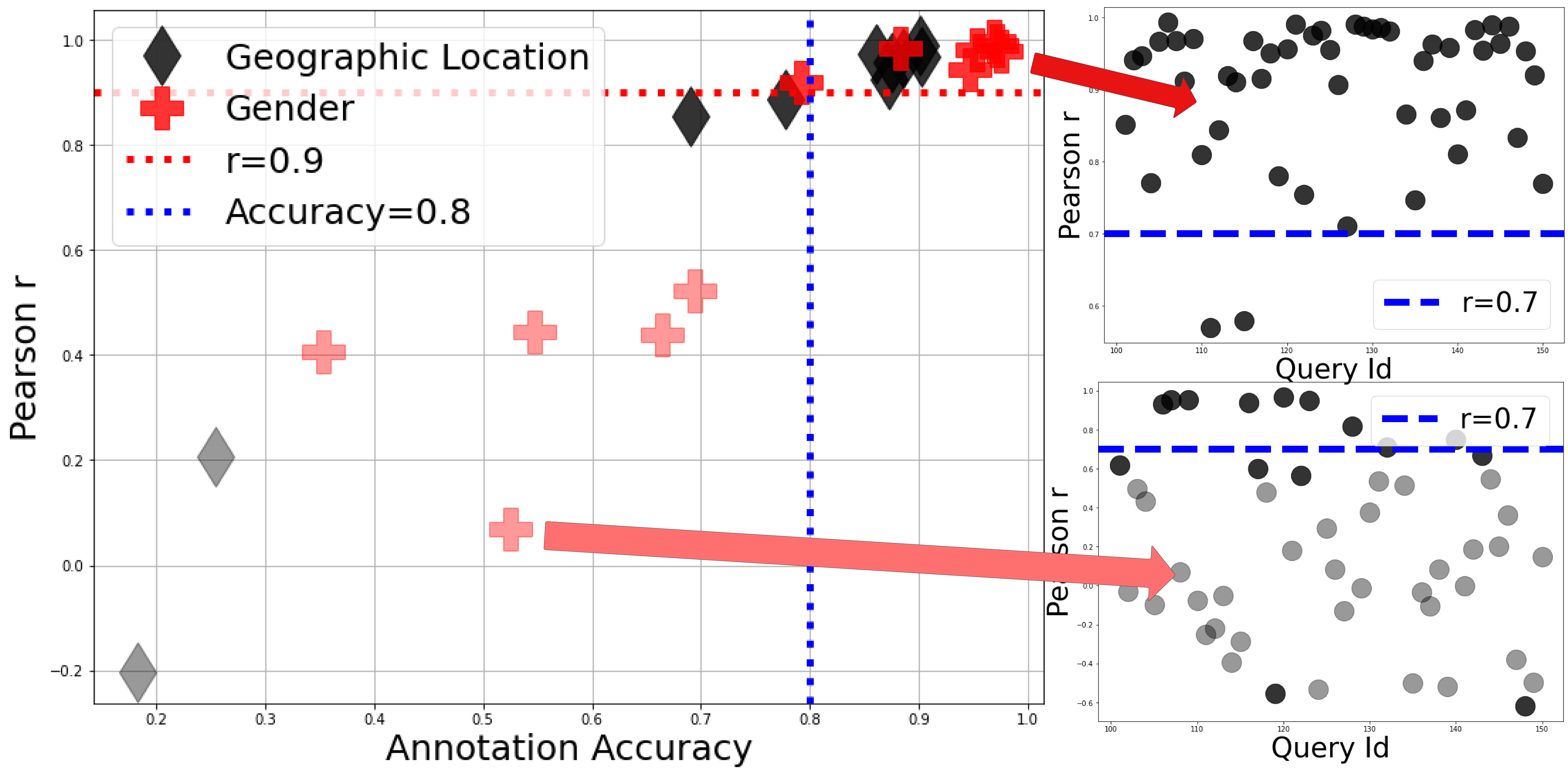}
    \caption{Annotation accuracy V.S. the correlation between evaluation metrics using human annotation and BERT-based annotation (TREC 2021). Shaded dots indicate $p > 0.05$. Two samples from the left plot with different annotation accuracy are selected to show query-level robustness.}
    \label{fig:samplesizegeo-r}
    \vspace{-5mm}
\end{figure}

So far, we know that to preserve the ability to differentiate the fairness of different systems, we need text classifier to be accurate, and minimal annotation errors will not degrade the ability. However, what are the impacts if the classifiers are not very precise, and how accurate do we need to be confident when replacing human annotations? To further explore the relationship between annotation accuracy and the corresponding evaluation correlation with the official fairness evaluation metrics, we would like to see the impact of annotation accuracy on the group fairness evaluation metrics. The first step is to obtain annotators with different annotation accuracy. According to Fig. \ref{fig:q-level-pear}, varying training sample size is one of the easiest ways to get different annotation models with different annotation accuracy. Hence, we trained the BERT-based annotation models with varying sample sizes and obtained several annotation models with different accuracy. The relation between the annotation accuracy and the effectiveness (Pearson r) of replacing human annotation with BERT-based annotation is plotted in Fig. \ref{fig:samplesizegeo-r}. As can be seen, generally, increasing annotation accuracy can not only improve system-level correlation to the official metrics but also improve query-level robustness. With an annotation accuracy above 0.8, using BERT-based annotation highly preserved the ability to differentiate fairness among different systems. Therefore, if group fairness evaluation focuses on the system level, minimal annotation errors could be ignored to save human efforts.

\vspace{-3mm}
\subsection{Generalizability of System Evaluation}
\vspace{-1mm}
GM annotation models can also be used for other complex evaluation metrics, such as evaluating a sequence of rankings. For example, the second task of the TREC fair ranking track 2021 evaluates fairness of sequence of rankings $\mathcal{L}_q$ by the expected exposure loss ($\text{EE-L}(\mathcal{L}_q) = || \gamma - \gamma^* ||$) \cite{diaz2020evaluating}, expected exposure disparity ($\text{EE-D}=||\gamma^*||_2^2$), and expected exposure relevance ($\text{EE-R}=2\gamma^T_{\pi}\gamma^*$). 

With GM annotation obtained from our BERT-based model, we compute the correlations between the TREC fair ranking track 2021 task 2's official evaluation metrics and those based on our annotation model based on 11 official submitted runs. We achieved Pearson correlation coefficients of 0.75, 0.98, and 0.79 for EE-L, EE-D, and EE-R, respectively, and all coefficient tests are statistically significant. Since the EE-D measures the inequality in exposure distribution across groups, and the BERT-based model has a similar performance across different groups for gender and geographic locations, we achieved a higher correlation with EE-D than the EE-R, which measures the agreement between exposure and relevance. Given the high correlation coefficients, especially for the pure fairness measure EE-D, our annotation model can also effectively replace human annotation for these fairness evaluation metrics.

\vspace{-4mm}
\section{Conclusion}
\vspace{-2mm}
Group membership, as one of the indispensable components in group fairness evaluation, requires massive human efforts to obtain. The sparsity of GM annotations limits the application of fairness evaluation and impedes fair ranking studies on general IR datasets. To overcome this, we compared four different language model-based text classifications for GM annotation. The BERT-based model achieved promising annotation accuracy with small-size training samples and less computational cost. Our query- and system-level evaluations confirmed the effectiveness and robustness of replacing human GM annotation with the BERT-based GM annotation. This opens a new direction to augment existing IR datasets for fairness evaluation and future fair-ranking studies. Even though LLMs have been used for mainstream NLP tasks and achieved impressive performance, they failed to outperform BERT for fairness GM annotation tasks as they were not designed for discriminative tasks. Moreover, according to our impact-oriented evaluation, when replacing human annotation with different annotators that have different annotation accuracy, minimal annotation errors will not degrade the fairness evaluation metrics. In the future, we would like to utilize the new annotation strategy to augment existing IR datasets for fairness studies, including fairness evaluation and fair ranking algorithms.

\vspace{-2mm}

\begin{credits}
\subsubsection{\ackname} This study is supported by the IFSA at the University of Delaware. We would like to thank the reviewers for their invaluable comments and suggestions.
\end{credits}
%
%
%
\bibliographystyle{splncs04}
\bibliography{ref}

\end{document}